\documentstyle[aps,prl,floats,graphicx,epsfig]{revtex}
\begin{document}
\draft
\twocolumn[\hsize\textwidth\columnwidth\hsize\csname @twocolumnfalse\endcsname

%
\title{Rotating vortex core: An instrument for detecting the
core excitations}
\author{N.B. Kopnin$^{1-3}$ and G.E. Volovik$^{1,3}$}
\address{
$^1$Landau Institute for Theoretical Physics, 117334 Moscow,
Russia\\
$^2$Laboratoire de Physique des Solides,
Universit\'e Paris-Sud, B{\^a}t 510, 91405 Orsay, France\\
$^3$Helsinki University of Technology, Low Temperature Laboratory,
P. O. Box 2200, FIN-02015 HUT, Finland }

\date{\today} \maketitle

\begin{abstract}

Effects of fermionic zero modes (bound states in a vortex core)
on the rotational dynamics of vortices with spontaneously broken
axisymmetry are
considered. The results are compared with the Helsinki NMR experiments
where the
vortex cores were driven to a fast rotation and torsional oscillations by
an NMR
r.f.\ field.  We predict a resonance NMR absorption on  localized states
at the external frequency  comparable with the interlevel distance which is
similar to the cyclotron Landau damping. The resonances can
resolve the localized levels in vortex cores. For a pure rotation of the
core, the  effect depends on the relative signs of the vortex
winding and of the core rotation; thus it is sensitive to the direction
of rotation of the container.  The similarity with fermionic zero modes
on fundamental strings, which simulate the thermodynamics of black holes,
is discussed.
\end{abstract}

\


\
]

\section{Introduction}

Low-frequency dynamics of vortices in Fermi superfluids and superconductors
is determined by low-energy excitations. If the energy spectrum of
fermionic quasiparticles
in the bulk superfluid/superconducting system has a finite gap, the relevant
low-energy quasiparticles occur only in vortex cores. The spectrum of
fermions bound to
the vortex core has an anomalous branch, which crosses zero as a function
of the angular
momentum. It was first found by Caroli, de Gennes and Matricon
\cite{Caroli} for
axisymmetric vortices in
$s$-wave superconductors. Effects of these quasiparticles on vortex
dynamics were considered
in a number of papers (see, for example, Refs.
\onlinecite{KopKr,KL}); some of the effects have the same origin as the
chiral anomaly in quantum field theory \cite{Bevan}.  New phenomena appear if
pairing occurs into a state with a different order-parameter symmetry.
For example, a new dynamics has
been recently predicted for vortices in $d$-wave superconductors:
Due to the localized excitations which are concentrated
near the gap nodes a finite dissipation exists even in the limit of a very
long relaxation time. This happens due to the Landau damping on
zero-sound collective modes in the 1D Fermi-liquid of
the localized states. \cite{KopninVolovik1997} The particular importance of
bound states near the gap nodes in $d$-wave superconductors has been already
demonstrated for the density of states \cite{KopninVolovik}: a
single-vortex density of states
was predicted to exhibit a singular $1/E$ energy dependence resulting in
a temperature independent vortex contribution to the specific heat.

In the present paper, we consider effects of bound states on dynamics of
nonaxisymmetric vortices which may exist in superfluids with unconventional
pairing. In particular, we consider vortices whose nonaxysymmetric cores
can rotate with respect to the heat bath.
This case is distinct from dynamics of (generally nonaxisymmetric) vortices
in $d$-wave superconductors
where however the orientation of vortex cores is fixed by the crystal
lattice, and
only a translational motion is possible. Such nonaxisymmetric vortices exist,
for example, in superfluid $^3$He-B where the superfluid environment and
the heat bath are
homogeneous and isotropic, while the vortex core exhibits spontaneous
anisotropy in the
crossectional plane. This spontaneous breaking of the continuous $U(1)$
symmetry (axial
symmetry)  of a single vortex core occurs below the vortex-core transition
\cite{Thuneberg,Salomaa} and is similar to the breaking of the
electromagnetic $U(1)$ symmetry in the core of a cosmic string
(the so-called superconducting string). \cite{Witten}

In $^3$He-B
vortex experiments, the nonaxisymmetric vortex cores were driven to fast
rotational oscillations plus rotation with a constant velocity under an
applied r.f.\ field. \cite{Kondo} In the present paper, we resolve the
existing controversy
concerning estimations of the effect of the core anisotropy on  vortex
dynamics \cite{Kondo}. We find that the main effects come from the fermionic
zero modes in the nonaxisymmetric vortex core. In particular, they lead to
resonances which should be observed at lower temperatures when the external
frequency of the core oscillations $\omega$ is a multiple of the distance
between the energy quantum levels in the core. Such resonances were first
predicted in Ref. \onlinecite{Kopnin/fr-dep-resp} for the translational
motion of
the axisymmetric vortices in superconductors,  but they were never observed
due to
a difficulty of exciting translational vortex oscillations with the  required
frequency.

We also consider the case when the core rotates around its axis with a
constant angular velocity $\Omega$. The frictional torque has an asymmetry
with respect to the relative signs of the vortex winding number and the
sense of core rotation; it thus can be detected by changing the direction of
one of the rotations. When $\Omega$ becomes equal to the interlevel spacing
gapless fermions appear; this has a close similarity to a $d$-wave
superconductor.

Finally, we discuss our results in connection with the similarity between
the fermionic zero modes in cores of vortices in condensed matter and
fermionic zero modes in cores of topological objects in cosmology (cosmic
strings, walls, monopoles, etc.). This similarity allows to model the process
of electroweak baryogenesis in the core of a cosmic string  using the
superfluid $^3$He-B. \cite{Bevan} We consider here another aspect of vortex
dynamics related to fermionic zero modes: namely, the  possibility to
simulate the event horizon and Hawking radiation by a rotating vortex core.
In this connection, the fermionic zero modes in a vortex core act exactly as
zero modes in fundamental strings in string description of black holes.

\section{HPD DRIVEN CORE OSCILLATIONS}

In Ref. \onlinecite{Kondo} the asymmetric cores of vortices
were coherently driven into a rotational and oscillatory motion using a
special and unique dynamical state with the coherently precessing
magnetization, known as the Homogeneously Precessing Domain (HPD).
This phase-coherent
precession is a magnetic counterpart of a superfluid coherent state and has the
similar off-diagonal long-range order, thus representing the magnetic
superfluidity.  The principles of HPD can be found in the review
\cite{Bunkov}.
The coherent precession has been used to study different properties of
superfluid
$^3$He-B. In particular, (i) vortices with an asymmetric core \cite{Kondo}
(analogs of the Witten superconducting strings \cite{Witten}) and (ii) vortices
formed at termination lines of solitons \cite{Kondo2} (analog of strings as
terminations of walls in cosmology
\cite{Review2}), have been detected using HPD.

In the HPD, the magnetization precesses with the
r.f.\ frequency $\omega_{{\rm RF}}$. The core is driven into motion due
to the dipole torque according to the following equation \cite{Kondo}:
\begin{equation}
f_{{\rm fr}}\dot\gamma=- T_D\sin 2(\omega_{{\rm RF}}t -\gamma)+
T_R \sin 2\gamma . \label{CoreEquation}
\end{equation}
Here $\gamma$ is the angle of the core anisotropy director ${\bf d}$ in the
plane transverse to the vortex axis, ${\bf d}=\hat x \cos\gamma + \hat y
\sin\gamma$ (see Fig.a); the dipole torque, $T_D$, and the friction parameter,
$f_{{\rm fr}}$, are nonzero if the core is anisotropic. The restoring
torque $T_R$
appears in a tilted magnetic field which determines a preferred orientation
for the vortex anisotropy director. Under the condition of the
experiment, the friction torque $\sim\omega_{{\rm RF}}f_{{\rm fr}}$ is large
compared with the driving dipole torque $T_D$. As a result, the core
performs small and fast oscillations superimposed on a slower drift
$\beta (t)$:
\begin{equation}
\gamma(t)= \gamma_\omega \cos 2(\omega _{\rm RF}t-\beta ) + \beta .
\label{CoreMotion}
\end{equation}
The effective drift velocity is nonzero for a comparatively small restoring
torque $T_R < T_D^2/2\omega _{\rm RF}f_{\rm fr}$. From Eq.(\ref{CoreEquation})
it follows that for zero restoring torque,  the drift angular velocity is
$\Omega
\equiv \dot
\beta =T_D^2/2f_{\rm fr}^2\omega _{\rm RF}$, i.e.,  it is much smaller than
the HPD
precession frequency. The amplitude of core  oscillations is $\gamma _\omega
=T_D/2f_{\rm fr}\omega _{\rm RF}$,

In principle, the core can perform a pure rotational motion, too. This
happens either for small friction $2\omega _{\rm RF}f_{\rm fr}<T_D$ when
$\Omega =\omega _{\rm RF}$, or if the core is twisted by the NMR r.f. field:
after the the external drive is switched off the core starts to untwist
through a free motion. \cite{Kondo} We consider here both pure rotation and
pure oscillations.

The dominating contribution to the friction comes from kinetics of fermions
in the rotating core which is thus the primary object of our investigation.
We find that Eq.(\ref{CoreEquation}) has a very limited applicability since
it does not contain the proper frequency dependence. We derive the correct
vortex dynamics which also includes the resonance absorption, i.e., the
Landau damping on the ``zero-sound-like'' modes in the system of localized
fermions in the vortex core.

\section{Effect of fermionic zero modes on core dynamics.}

{\it Quasiparticles bound to the vortex}. The low-energy spectrum of
Caroli-de-Gennes-Matricon quasiparticles around a vortex contains an
anomalous branch of fermionic zero modes. Excitations on this branch are
characterized by two canonically conjugated variables, the angle $\alpha$
of the
transverse linear momentum ${\bf p_\perp}$ of the quasiparticle, say with
respect
to the core anisotropy direction ${\bf d}$ (see Fig.a), and the angular
momentum
$Q$. For superconductors with the coherence length
$\xi$ much larger than the inverse Fermi momentum, $p_F\xi \gg 1$, the electron
wavelength is short compared with the core size, and the quasiclassical
approximation is relevant. In the quasiclassical approximation, the variables
$\alpha $ and $Q$ commute and the spectrum has the form
\begin{equation}
E(Q,\alpha )=-\omega _{0}(\alpha )[Q-Q_{0}(\alpha )].  \label{CdGM/spectrum}
\end{equation}
The quasiclassical angular momentum $Q$ is a continuous variable; thus the
anomalous branch crosses
zero as a function of $Q$ at $Q=Q_{0}(\alpha )$. In particular, one has
$Q_{0}=0$
when the vortex-core structure is completely symmetric. $Q_{0}$ becomes finite
if parity is violated. This was found by Schopohl \cite{Schopohl}
for a nonaxisymmetric vortex in $^{3}$He-B; two anomalous branches, which cross
zero at two antisymmetric values, $Q=\pm Q_0(\alpha)$, are schematically
shown on
Fig.b. The magnitude of
$Q_{0}(\alpha )
$ in continuous vortices in $^{3}$He-A is determined by the core size of the
vortex $|Q_{0}(\alpha )|\sim p_{F}R_{{\rm core}}$ \cite
{KopninPhysica,OrbitalMomentum}. For $^{3}$He-B, where the core radius is of
order of coherence length $\xi $, this estimation gives a large value of $
Q_{0}\sim p_{F}\xi $, in accordance with the unpublished calculations by
Schopohl for the nonaxisymmetric vortex with broken parity, the so called
$v$-vortex. \cite{Schopohl}

\begin{figure}[!!!t]
\begin{center}
\leavevmode
\epsfig{file=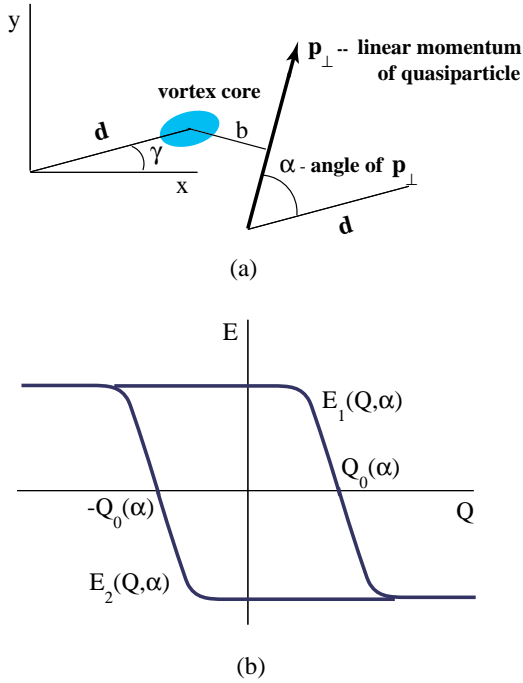,width=0.8\linewidth}
\caption[trajectory]
    {(a) Trajectory of the quasiparticle near the asymmetric vortex core.
Angle $\gamma$ shows the orientation of the core anisotropy vector ${\bf
d}$; angle
$\alpha$ shows the direction of the transverse linear momentum
${\bf p_\perp}$ of the quasiparticle with respect to ${\bf d}$; and $b$ is the
impact parameter. (b) Quantization of the fast motion along the trajectory
gives
two branches of low energy spectrum
$E_{1,2}(b,\alpha)$ in terms of the rest (slow) variables, $b$ and $\alpha$.
Instead of the impact parameter it is better to use  the continuous angular
momentum
$Q=b|{\bf p_\perp}|$. As a function of $Q$ the spectrum $E_{1,2}(Q,\alpha)$
crosses zero at two antisymmetric values of
$Q$, which depend on
$\alpha$: $Q=\pm Q_0(\alpha)$.  At the next stage the  true quantum mechanical
levels  are obtained by quantization of slow motion using the Bohr-Sommerfeld
rule for the canonically conjugated variables
$Q$ and $\alpha$: $\oint d\alpha Q(\alpha,E)=2\pi n \hbar$. This gives the
equidistant levels $E=nE_0$.}
\label{trajectory}
\end{center}
\end{figure}

{\it Kinetic equation} for the distribution function $f(t,\alpha ,Q)$ of
fermions on the anomalous branch for a system characterized by canonically
conjugated variables has a conventional form \cite{Stone}:
\begin{equation}
{\frac{\partial f}{\partial t}}+{\frac{\partial f}{\partial \alpha }}{\frac{
\partial E}{\partial Q}}-{\frac{\partial E}{\partial \alpha }}{\frac{
\partial f}{\partial Q}}=-{\frac{f-f_{0}}{\tau }}.  \label{KE}
\end{equation}
Here $\partial E/\partial Q=-\omega _{0}(\alpha )$; $f_{0}$ is the
equilibrium Fermi distribution function.

If the vortex core rotates the quasiparticle energy depends on time
according to $E(Q,\alpha ,t)=E(Q,\alpha -\gamma (t))$. After transformation
$\alpha -\gamma (t) \rightarrow \alpha$ one obtains the following kinetic
equation:
\begin{equation}
{\frac{\partial f}{\partial t}}-(\omega _{0}+\dot{\gamma}){\frac{\partial f}{
\partial \alpha }}-{\frac{\partial E}{\partial \alpha }}{\frac{\partial f}{
\partial Q}}=-{\frac{f-f_{0}}{\tau }}.  \label{KEshifted}
\end{equation}

One can write
\begin{equation}
f_{1}=f-f_{0}=(df_{0}/dE)\chi .  \label{f1}
\end{equation}
If $\dot{\gamma}=\dot{\gamma}_{\omega }e^{-i\omega t}=-i\omega \gamma
_{\omega }e^{-i\omega t}$ is a perturbation with a frequency
$\omega =2\omega _{\rm RF}$, one has
\begin{equation}
\left( -i\omega +{\frac{1}{\tau }}\right) \chi -\omega _{0}\frac{\partial
\chi }{\partial \alpha }-\frac{\partial E}{\partial \alpha }\frac{\partial
\chi }{\partial Q}=\dot{\gamma}_{\omega }\frac{\partial E}{\partial \alpha }.
\label{KEfor(chi)}
\end{equation}

{\it Friction torque} on a rotating core from fermions on the bound states
in the core is
\begin{eqnarray}
T_{{\rm f}}(\omega)= -\int \frac{d p_z}{2\pi } \int {\frac{d \alpha dQ}{2\pi}
}~\frac{\partial Q}{\partial t} \chi {\frac{\partial f_0}{\partial E}}
\nonumber \\
=\int \frac{d p_z}{2\pi } \int {\frac{d \alpha dQ}{2\pi}} {\frac{\partial E}{
\partial \alpha}}{\frac{df_0}{dE}}\chi .  \label{Torque}
\end{eqnarray}

{\it The eigen-states} or zero-sound-like collective modes in the
Fermi-liquid of localized states are obtained as solutions of the homogeneous
equation, the Eq.(\ref{KEfor(chi)}) without the driving term on the right hand
side:
\begin{eqnarray}
\chi _{0}(\alpha ,Q,\omega ) &=&\Lambda [E(\alpha ,Q)]e^{F(\alpha ,\omega )},
\label{SolutionOfUniformEquation} \\
F(\alpha ,\omega ) &=&\int_{0}^{\alpha }d\alpha ^{\prime }g(\alpha ^{\prime
},\omega )~,~g(\alpha ,\omega )=\frac{-i\omega +1/\tau }{\omega _{0}(\alpha )
}  \label{F}
\end{eqnarray}
where $\Lambda (E)$ is an arbitrary function of $E$. The eigenvalues are
obtained from the requirement of periodicity $\chi (\alpha ,Q)=\chi (\alpha
+2\pi ,Q)$:
\begin{equation}
\int_{0}^{2\pi }d\alpha ~g(\alpha ,\omega )=2\pi ni,  \label{Quantization}
\end{equation}
where $n$ is an integer. In the collisionless regime, this gives
eigenfrequencies of the collective modes:
\begin{equation}
\omega =nE_{0}~,~{\frac{1}{E_{0}}}=\int_{0}^{2\pi }{\frac{d\alpha }{2\pi }}{
\frac{1}{\omega _{0}(\alpha )}}\equiv \left\langle {\frac{1}{\omega _{0}}}
\right\rangle ,  \label{EigenFrequencies}
\end{equation}
Here $E_{0}(p_{z})$ is an exact minigap as distinct from the quasiclassical
minigap $\omega _{0}(p_{z})$ (compare with Ref.
\onlinecite{KopninVolovik1997}). The Landau damping on these discrete
frequencies occurs even in the collisionless regime. It is completely
analogous to the Landau cyclotron damping. Let us consider this in more
detail.

{\it General solution} of the Eq. (\ref{KEfor(chi)}) can be found from the
following consideration. Since both the energy $E(\alpha ,Q)=-\omega
_{0}(Q-Q_{0})$ and the r.h.s. of Eq.(\ref{KEfor(chi)}) are binomials of $Q$,
a general solution can also be represented as a binomial in $Q$, and thus as a
binomial in $E$:
\begin{equation}
\chi (\alpha ,Q,\omega )=[a(\alpha ,\omega )E(\alpha ,Q)+b(\alpha ,\omega
)]e^{F(\alpha ,\omega )}.  \label{GeneralSolutionNonuniformEq}
\end{equation}
The functions $a$ and $b$
satisfy the equations
\begin{equation}
\frac{\partial a}{\partial \alpha }=-\frac{\dot{\gamma}_{\omega }}{\omega
_{0}^{2}}\frac{\partial \omega _{0}}{\partial \alpha }e^{-F(\alpha ,\omega
)},\frac{\partial b}{\partial \alpha }=-\dot{\gamma}_{\omega }\frac{\partial
Q_{0}}{\partial \alpha }e^{-F(\alpha ,\omega )}.  \label{EqFor-b}
\end{equation}

At small $T$ the linear-in-$E$ term in Eq. (\ref{GeneralSolutionNonuniformEq}
) is small and thus only the equation for $b(\alpha ,\omega )$ is important.
The solution for $b$ consistent with the periodicity of $\chi $ gives for $
E\to 0$
\begin{eqnarray}
\chi (\alpha ,\omega ) &=&-\dot{\gamma}_{\omega }e^{F(\alpha ,\omega )}
\Biggl[ \int_{0}^{\alpha }d\alpha ^{\prime }\frac{\partial Q_{0}}{\partial
\alpha ^{\prime }}e^{-F(\alpha ^{\prime },\omega )}  \nonumber \\
&&+{\frac{1}{{e^{-F(\omega )}-1}}}\int_{0}^{2\pi }d\alpha \frac{\partial
Q_{0}}{\partial \alpha }e^{-F(\alpha ,\omega )}\Biggr] ,
\label{ChiSolutionE=0}
\end{eqnarray}
where $F(\omega )\equiv F(2\pi ,\omega )$.

{\it Landau cyclotron damping.} In the collisionless limit and close to a
pole at $\omega =nE_{0}$ one has
\begin{eqnarray}
\chi (\alpha ,\omega  &\approx &nE_{0})=-{\frac{\dot{\gamma}_{\omega }}{2\pi
i}}~\frac{E_{0}}{\omega -nE_{0}+i/\tau }  \nonumber \\
&&\times e^{F(\alpha ,nE_{0})}\int_{0}^{2\pi }d\alpha \frac{\partial Q_{0}}{
\partial \alpha }e^{-F(\alpha ,nE_{0})}.  \label{ChiCloseToPoles}
\end{eqnarray}
The torque on the core is
\begin{equation}
T_{{\rm f}}(\omega )=-\dot{\gamma}_{\omega }\int \frac{dp_{z}}{2\pi }
G(\omega ,p_{z})\equiv -\dot{\gamma}f_{{\rm fr}}(\omega )  \nonumber
\end{equation}

Close to the resonance one has the response
\begin{eqnarray}
G(\omega,p_z)= \frac{i E_0}{ \omega - nE_0 +i/\tau}  \nonumber \\
\times \left<\frac{\partial Q_0}{\partial \alpha}e^{F(\alpha,nE_0)}\right>
\left< \frac{\partial Q_0}{\partial \alpha}e^{-F(\alpha,nE_0)}\right> .
\label{ResponseResonance}
\end{eqnarray}

Let us consider the resonance absorption in the simple model in which $
\omega_0$ does not depend on $\alpha$ while the core anisotropy is
represented only by the function $Q_0(\alpha)$. This model is close to what
was found by Schopohl \cite{Schopohl}. In the simple model of constant $
\omega_0$, one has
\[
G(\omega,p_z)= \frac{iE_0 n^2|Q_{0n}|^2 }{\omega - nE_0 +i/\tau }
~,~Q_{0n}=\int {\frac{d\alpha}{2\pi}} Q_0(\alpha) e^{in\alpha},
\]
where $Q_{0n}$ is the $n$-th harmonic of the $Q_0(\alpha)$.

The $p_z$ dispersion of the exact minigap $E_0(p_z)$ results in formation of
an absorption band \cite{Kopnin/fr-dep-resp}. If the frequency $\omega$
falls within the band, the poles in $G(\omega,p_z)$ give rise to the Landau
damping even in the completely collisionless limit: the friction coefficient
is
\begin{equation}
{\rm Re} f_{{\rm fr}} (\omega)=\sum_{n>0} E_0(p_{zn}) n |Q_{0n}(p_{zn})|^2
\left[ \left|{\frac{dE_0}{dp_z}}\right|_{p_{zn}} \right]^{-1}
\label{CollisionlessRealG}
\end{equation}
where $p_{zn}$ is the momentum for which the resonance condition, $
nE_0(p_{zn})=\omega$, takes place.

{\it NMR absorption: Collisionless regime.} Eq. (\ref{CoreEquation}) is
modified due to the frequency dispersion of the response function:
\begin{equation}
f_{{\rm fr}}(\omega)\dot\gamma_\omega=- T_D ,  \label{CoreEquationModified}
\end{equation}
which gives the NMR resonant absorption
\begin{equation}
P = {\frac{1}{2}} {\rm Re} f_{{\rm fr}}(\omega)|\dot\gamma_\omega|^2= {\frac{
1}{2}} T_D^2 {\rm Re}\, f_{{\rm fr}}^{-1}(\omega) .
\label{NMRAbsorptionGeneral}
\end{equation}
In the collisionless regime, Eq. (\ref{CollisionlessRealG}) can be used. The
resonant absorption starts with a sharp minimum in the background
absorption. Indeed, if $\omega$ is at the edge of the band $nE_0(p_z)$ where
$dE_0/ dp_z=0$ the absorption proportional to $[{\rm Re}\, f_{{\rm fr}
}(\omega)]/|f_{{\rm fr}}(\omega)|^2$ decreases sharply due to an infinitely
large real part of $f_{{\rm fr}}(\omega )$. It goes exactly to zero in the
ideal collisionless limit. Observation of these ``van Hove'' zeroes would
resolve the minigap in NMR experiments.

{\it NMR absorption: Hydrodynamic limit}. In the hydrodynamic limit, $\omega
_0\tau \ll 1$, we have $F(\omega )\gg 1$, and the first term in Eq. (\ref
{ChiSolutionE=0}) dominates. It results in a local expression
\begin{equation}
\chi =\dot \gamma_\omega \frac{\partial Q_0}{\partial \alpha }\frac{\omega _0
} {-i\omega +1/\tau}  \label{ChiSolutionHydro}
\end{equation}
The friction parameter which enters the NMR absorption in Eq.(\ref
{NMRAbsorptionGeneral}) becomes
\begin{equation}
{\rm Re}\, f_{{\rm fr}}^{-1}(\omega)={\frac{1}{\tau}}\left[\frac{p_F}{\pi}
\left<\omega_0 \left({\frac{\partial Q_0}{\partial \alpha}}\right)^2\right>
_F\right]^{-1} .  \label{FrictionOscillatory}
\end{equation}
where $\left< \ldots \right>_F$ is the average over the Fermi surface.

The experiment of Ref. \onlinecite{Kondo}  has been carried out in the
temperature range $0.48 T_c< T < 0.6 T_c$ while, according to Ref.
\onlinecite{Bevan}, the hydrodynamic regime $\tau \omega_0 < 1$ occurs only
above $T\sim 0.6 T_c$. Let us compare, however, the theory and experiment at
$T= 0.6 T_c$. The experimental value of the friction parameter in Eq.(\ref
{CoreEquation}), $f_{{\rm fr}}\sim 5\times 10^{-18}$ erg s/cm at $T= 0.6 T_c$
\cite{Kondo}, can be obtained if one takes $\sqrt {<(\partial Q_0/\partial
\alpha )^2>} \sim 10 $. This large magnitude of the derivative $\partial
Q_0/\partial \alpha $ is not inconsistent with the Schopohl result: $Q_0
\sim p_F\xi$. However, since $T$ is not small in the experiment,
the $T^2$ term which results from the omitted energy-dependent terms
in Eq. (\ref{ChiSolutionE=0}) can be important and can give a comparable
contribution to $f_{{\rm fr}}$. The temperature is to be decreased to
resolve the value of $Q_0$.

\section{Uniform core rotation.}

If the vortex core rotates with a constant angular velocity $\Omega $ the
asymptotic solution is time independent and obeys the equation
\begin{equation}
\frac{1}{\tau }\chi -(\omega _{0}+\Omega ){\frac{\partial \chi }{\partial
\alpha }}-{\frac{\partial E}{\partial \alpha }}{\frac{\partial \chi }{
\partial Q}}=\Omega {\frac{\partial E}{\partial \alpha }}.
\label{KEstationary}
\end{equation}

The case of a slow rotation is equivalent to Eq. (\ref{ChiSolutionHydro})
with $\omega =0$ and $\dot \gamma =\Omega$. In the hydrodynamic limit, we
thus have Eq. (\ref{FrictionOscillatory}).

The case of a fast rotation $\Omega \sim \omega _{0}$ can be considered as a
linear responce only in the limit of a small anisotropy, $\partial \omega
_{0}/\partial \alpha \ll \omega _{0}$ and $\partial Q_{0}/\partial \alpha
\ll Q_{0}$. Using the perturbation theory in the anisotropy parameters, we
can solve the kinetic equations for both high and small relaxation rate.
Using the ansatz of Eq. (\ref{GeneralSolutionNonuniformEq}), we find the
periodic solution
\begin{equation}
\chi =\int_{0}^{2\pi }d\alpha ^{\prime }\left[ {\frac{\Theta (\alpha -\alpha
^{\prime })}{e^{\Phi }-1}}+{\frac{\Theta (\alpha ^{\prime }-\alpha )}{
1-e^{-\Phi }}}\right] b(\alpha ^{\prime })e^{\Phi (\alpha -\alpha ^{^{\prime
}})}  \label{SolutionFor(v)}
\end{equation}
where
\begin{eqnarray}
b(\alpha ) &=&{\frac{\Omega \omega _{0}}{\omega _{0}+\Omega }}{\frac{
\partial Q_{0}}{\partial \alpha }},  \label{b} \\
\Phi (\alpha ) &=&{\frac{\alpha }{\tau (\omega _{0}+\Omega )}}~,~\Phi \equiv
\Phi (2\pi ).  \label{Phi}
\end{eqnarray}

For $T\to 0$ when $df_{0}/dE=-\delta (Q-Q_{0})/\omega _{0}$ and $dE/d\alpha
=\omega _{0}dQ_{0}/d\alpha $, one obtains the response function
\begin{eqnarray}
G(0) ={\frac{\omega _{0}}{\omega _{0}+\Omega }}\int_{0}^{2\pi }{\frac{
d\alpha d\alpha ^{\prime }}{2\pi }}  \nonumber \\
\times \left[ {\frac{\Theta (\alpha -\alpha ^{\prime })}{e^{\Phi }-1}}+{
\frac{\Theta (\alpha ^{\prime }-\alpha )}{1-e^{-\Phi }}}\right] {\frac{
\partial Q_{0}}{\partial \alpha }}{\frac{\partial Q_{0}}{\partial \alpha
^{\prime }}}e^{\Phi (\alpha -\alpha ^{^{\prime }})}.  \label{GeneralFriction}
\end{eqnarray}

{\it Resonance}. Eq. (\ref{GeneralFriction}) essentially depends on the
sense of the rotation of the vessel, which determines the relative signs of $
\omega _{0}$ and $\Omega $. The most interesting situation occurs when $
\omega _{0}$ and $\Omega $ have opposite signs. In this case the resonance
absorption should occur when $-\Omega $ is close to $\omega _{0}$. In the
hydrodynamic regime, $|\omega _{0}+\Omega |\ll 1/\tau $, the frictional
torque coincides with Eq.(\ref{FrictionOscillatory}). In the intermediate
regime, $|\omega _{0}+\Omega |\gg 1/\tau \gg |\omega _{0}-<\omega >|$, one
obtains
\begin{equation}
f_{{\rm fr}}=\frac{p_{F}}{\pi }\left\langle {\frac{\omega _{0}}{(\omega
_{0}+\Omega )^{2}}}~{\frac{\left\langle Q_{0}^{2}\right\rangle -\left\langle
Q_{0}\right\rangle ^{2}}{\tau }}\right\rangle _{F}.
\label{FrictionIntermediate}
\end{equation}
Using Eq. (\ref{FrictionIntermediate}) and Eq. (\ref{FrictionOscillatory}),
one can approximate the resonance absorption by
\begin{equation}
f_{{\rm fr}}=p_{F}\left\langle \omega _{0}\delta (\omega _{0}+\Omega
)~\left[ \left\langle Q_{0}^{2}\right\rangle -\left\langle
Q_{0}\right\rangle ^{2}\right] \right\rangle _{F},  \label{Resonance}
\end{equation}

If, for the proper sense of rotation, the function $\omega
_{0}(p_{z})+\Omega$   crosses zero at some
$ p_{z0}$, the friction remains finite even in the limit $\tau \gg 1/\omega
_{0}$ (however, $\tau $ has to be small compared to the inverse dispersion):
\begin{eqnarray}
f_{{\rm fr}} &=&\int \frac{dp_{z}}{2}\omega _{0}\delta (\omega _{0}+\Omega
)~\left[ \left\langle Q_{0}^{2}\right\rangle -\left\langle
Q_{0}\right\rangle ^{2}\right]   \nonumber \\
&=&\frac{\omega _{0}}{|d\omega _{0}/dp_{z}|_{p_{z0}}}~\left[ \left\langle
Q_{0}^{2}\right\rangle -\left\langle Q_{0}\right\rangle ^{2}\right] .
\label{FrictionTauInfty}
\end{eqnarray}

In the simple-core model where $\omega _{0}$ is independent of $\alpha $ and
$Q_{0}$ has only one harmonic,
\begin{equation}
Q_{0}(\alpha )=\pm g\cos \alpha ,  \label{ModelCore}
\end{equation}
the solution is
\begin{equation}
\chi =A\sin \alpha +B\cos \alpha ~,~A=-{\frac{\omega _{0}\Omega \tau g}{
1+(\omega _{0}+\Omega )^{2}\tau ^{2}}}.  \label{SolutionSinus}
\end{equation}
The factor $B$ is not important since it does not contribute to the friction
torque. The friction torque has a resonance behavior
\begin{equation}
G(p_{z})=\frac{g^{2}}{2}{\frac{\omega _{0}\tau }{1+(\omega _{0}+\Omega
)^{2}\tau ^{2}}}~.  \label{FrictionSinus}
\end{equation}
in agreement with both Eq. (\ref{FrictionIntermediate}) for collisionless
regime and Eq. (\ref{FrictionOscillatory}) for the hydrodynamic regime.

\section{Discussion}

The interesting problems to be discussed in the future are related to the
resonance behavior, which occurs when the edge of the minigap band $\omega
_{0}(\alpha )$ as a function of $\alpha $ crosses the angular velocity $
\Omega $ of the unifrom rotation. If the bottom of the band is chosen at $
\alpha =0$ (and also at $\alpha =\pi $ due to symmetry of the core) one has
near the bottom
\[
\omega _{0}(\alpha )+\Omega \approx \omega _{0}(0)+\Omega +{\frac{1}{2}}
\beta \alpha ^{2}~,~\beta =d^{2}\omega _{0}(0)/d\alpha ^{2}.
\]
At the moment when $\Omega $ touches the bottom $\omega _{0}(0)$, the
shifted minigap $\omega _{0}(\alpha )+\Omega $ has nodes of a second order.
The same situation occurs for non-rotating vortices in the $d$-wave
superconductors. The higher-order gap nodes have serious consequences for
the vortex thermodynamics \cite{d-waveVortex,KopninVolovik} and dynamics
\cite{KopninVolovik1997}. In particular, the dissipation occurs even in the
superclean limit.

Let us start with a small positive $\omega _{0}(0)+\Omega $ and then
continue it analytically to a negative value. We are interested in the
Bose-factor in Eq.(\ref{SolutionFor(v)}) in the collisionless limit $\tau
=\infty $ with a finite frequency $\omega $. Substituting $1/\tau $ with $
-i\omega $, we find for $0<\omega _{0}(0)+\Omega \ll \omega _{0}(0)$
\begin{equation}
\Phi \approx -2i\omega \int_{-\pi /2}^{\pi /2}{\frac{d\alpha }{\omega
_{0}(0)+\Omega +{\frac{1}{2}}\beta \alpha ^{2}}}\approx -{\frac{2\pi i\omega
\sqrt{2}}{\sqrt{\beta (\omega _{0}(0)+\Omega )}}}.  \label{PhiBeforeCrossing}
\end{equation}
Here we took into account that the integral is concentrated in the vicinity
of $\alpha =0$ and also in the vicinity of $\alpha =\pi $. The analytical
continuation of Eq. (\ref{PhiBeforeCrossing}) to a small negative $\omega
_{0}(0)+\Omega $ gives a real quantity
\begin{equation}
\Phi \approx -\frac{2\pi \omega \sqrt{2}}{\sqrt{\beta \left| \Omega +\omega
_{0}(0)\right| }}=-4\pi \frac{\omega }{\left| \partial \omega _{0}/\partial
\alpha \right| _{\omega _{0}(\alpha )=-\Omega }}.  \label{PhiAfterCrossing}
\end{equation}
The Bose-factor in Eq. (\ref{SolutionFor(v)}) thus corresponds to an
equilibrium distribution with an effective temperature determined by the
derivative of the minigap $\omega _{0}(\alpha )$ at the point where $\omega
_{0}(\alpha )=-\Omega $:
\begin{equation}
\frac{1}{e^{\omega /T_{{\rm eff}}}-1}~,~T_{{\rm eff}}=\frac{1}{4\pi }\left|
\left( {\frac{\partial \omega _{0}}{\partial \alpha }}\right) _{\omega
_{0}(\alpha )=-\Omega }\right| .  \label{EffectiveT}
\end{equation}

This has a close relation with the event horizon problem \cite{Jacobson}.
Recently there was a profound progress in understanding of the
Bekenstein-Hawking \cite{Bekenstein,Hawking} black hole thermodynamics using
the fundamental strings (see recent review paper by Horowitz
\cite{Horowitz}). The
statistical origin of the black hole thermodynamics is related to fermion and
boson zero modes of strings. Excitations having the lowest energy are confined
to a circle and are chiral: they are, say, right-moving along the circle.
Fermions
localized in the core of a quantized vortex in a superconductor or in a
Fermi-superfluid of the type of superfluid $^3$He have similar
properties\cite{Caroli}.  The model of a vortex core where fermions live on a
closed line around the vortex axis in real space was discussed in Ref.
\onlinecite{Volovik/edge}. The lowest energy branch (the fermionic zero mode in
the vortex) represents  1 + 1 massless fermions, which are left-moving
(right-moving) around the  axis of
the vortex, if the vortex has a positive (negative) winding number. The
axial anomaly caused by the fermion zero modes during motion of vortices
(strings) in $^3$He and superconductors \cite{KopninVolovik1995} has
experimentally observable consequences for the vortex dynamics. \cite{Bevan}

If the canonically conjugated variables on the circle, i.e., the angle $
\alpha $ and the angular momentum $Q$, are represented as the coordinate $
x=\alpha $ and the momentum $p=Q$, the spectrum of fermionic zero modes of
the type of Eq. (\ref{CdGM/spectrum}) describes a massless relativistic
particle left-moving along the circle,
\begin{equation}
E(p,x)=-c(x)(p-A(x)).  \label{E(x)}
\end{equation}
Here $c(x)=\omega _{0}(x)$ plays the part of the speed of light, and $
Q_{0}(\alpha )=A(x)$ is the vector potential. This spectrum is universal,
though the details, i.e., the functions $c(x)=\omega _{0}(\alpha )$ and $
A(x)=Q_{0}(\alpha )$, depend on a particular vortex structure in a
particular superfluid/superconductor.

The gas of these 1 + 1 massless chiral fermions has the required statistical
properties. For example, if we consider the axisymmetric core, the
quasiparticle enrgy spectrum in terms of the angular momentum quantum
number $Q$ is
$E=-\omega_0 Q$. For such spectrum the number of quantum states  with the fixed
total angular momentum ${\cal Q}$ is \cite{Wald}
\begin{equation}
e^{S}~~,~~S=2\pi\sqrt{{\cal Q}/12}~.  \label{Entropy}
\end{equation}
The same happens also for the nonaxisymmetric vortex, since its energy
spectrum is also linear in quantum number, $E=nE_0$ according to
Eq.(\ref{EigenFrequencies}). This means that if the vortex has the nonzero
momentum
${\cal Q}$ (this happens for example if two identical vortices rotate about
their
center of mass or if the nonaxisymmetric core rotates) the  fermionic entropy
$S$ of the core appears to be nonzero even at $T=0$. This violation of the
"Nernst
theorem"  is similar to that in the black hole.
\cite {Horowitz,Wald} Also the discrete spectrum of the radiation by the
rotating
core, with frequencies which are integral multiple of $E_0$ in
Eq.(\ref{EigenFrequencies}), is similar to that by the quantum black
hole.\cite{Bekenstein97}

In the classical limit the rotation of the nonaxisymmetric vortex can lead
to the
``event horizon''. This happens when $| \Omega | >\omega _{0}(0)$
and the ``speed of light'', $ c(x)=\omega _{0}(\alpha ) + \Omega  $,
crosses zero at some values of $x$. The analogue of the Hawking temperature is
to be played by the spatial derivative of the speed of light at horizon,
$T_{\rm Hawking}=(1/2\pi )|(dc/dx)_{{\rm hor}}|=|(d\omega _{0}/d\alpha
)_{\omega _{0}(\alpha )=-\Omega }|$. This is two times larger than the
effective temperature in Eq.(\ref{EffectiveT}) probably because of two
``black holes'' in the core (close to $\alpha =0$ and close to $\alpha =\pi $).

To summarize, we predict a resonant absorption on localized levels in an
asymmetric vortex core performing rotational oscillations which are similar
to the cyclotron Landau damping. For a pure rotation of the core around its
axis with a constant angular velocity $\Omega$, the frictional torque has an
asymmetry with respect to the relative signs of the vortex winding number
and the sense of core rotation.

\acknowledgements

GEV thanks Ted Jacobson for the collaboration on the horizon problem. This
work was supported by the Russian Foundation for Fundamental Research grant
No. 96-02-16072 and by the program ``Statistical Physics'' of the Ministry of
Science of Russia. NBK also acknowledges the support by the Swiss National
Foundation cooperation grant 7SUP J048531 and by the INTAS grant 96-0610.
GEV acknowledges the support by the European Science Foundation.


\begin{references}
\bibitem{Caroli}  C. Caroli, P. G. de Gennes, and J. Matricon, Phys. Lett.
{\bf 9}, 307 (1964).

\bibitem{KopKr}  N. B. Kopnin and V. E. Kravtsov, Pis'ma Zh. Eksp. Teor.
Fiz. {\bf 23}, 631 (1976) [JETP Lett. {\bf 23}, 578 (1976)].

\bibitem{KL}  N. B. Kopnin and A. V. Lopatin, Phys. Rev. B {\bf 51}, 15291
(1995).

\bibitem{Bevan}  T. D. C. Bevan, A. J. Manninen, J. B. Cook, J. R. Hook, H.
E. Hall, T. Vachaspati and G. E. Volovik, Nature, {\bf 386}, 689 (1997).

\bibitem{KopninVolovik1997}  N. B. Kopnin and G. E. Volovik,
Phys. Rev. Lett. {\bf 79}, 1377 (1997).

\bibitem{KopninVolovik}  N. B. Kopnin and G. E. Volovik, Pis'ma ZhETF, {\bf
64}, 641 (1996); [JETP Lett., {\bf 64}, 690 (1996)].

\bibitem{Thuneberg}  E.V. Thuneberg, Phys. Rev. Lett. {\bf 56}, 359 (1986).

\bibitem{Salomaa}  M.M. Salomaa and G.E. Volovik, Phys. Rev. Lett. {\bf 56},
363 (1986).

\bibitem{Witten}  E. Witten, Nucl. Phys. {\bf B~ 249}, 557 (1985).

\bibitem{Kondo}  Y. Kondo, J. S. Korhonen, M. Krusius, V. V. Dmitriev, Yu.
M. Mukharskiy, E. B. Sonin and G. E. Volovik, Phys. Rev. Lett. {\bf 67}, 81
(1991).

\bibitem{Kopnin/fr-dep-resp}  N. B. Kopnin, Pis'ma Zh. Eksp. Teor. Fiz. {\bf
27}, 417 (1978) [JETP Lett. {\bf 27}, 390 (1978)].

\bibitem{Bunkov} Yu.M. Bunkov, in: {\it  Progress in Low Temperature Physics},
{\bf 14}, pp. 69-158, ed. W. Halperin, Elsevier, (1995).

\bibitem{Kondo2} Y. Kondo,  J.S. Korhonen, M. Krusius, V.V. Dmitriev,
  E.V. Thuneberg and  G.E. Volovik,    Phys. Rev. Lett. {\bf 68}, 3331 (1992).

\bibitem{Review2} A. Vilenkin, E. P. S.Shellard,
{\it Cosmic Strings and Other Topological Defects} (Cambridge University
Press,1994);
M. B. Hindmarsh, T. W. B. Kibble, Rep. Prog. Phys. {\bf 58}, 477 (1995).

\bibitem{Schopohl}  N. Schopohl, unpublished.

\bibitem{KopninPhysica}  N. B. Kopnin, Physica, {\bf B~210}, 267 (1995).

\bibitem{OrbitalMomentum}  G. E. Volovik, Pis'ma ZhETF {\bf 61 }, 935 (1995)
[JETP Lett. {\bf 61}, 958 (1995)].

\bibitem{Stone}  M. Stone, Phys. Rev. {\bf B} 54, 13222 (1996).

\bibitem{d-waveVortex}  G. E. Volovik, Pis'ma ZhETF, {\bf 58}, 457 (1993);
[JETP Lett., {\bf 58}, 469 (1993)].

\bibitem{Jacobson}  T. Jacobson and G. E. Volovik, to be published.

\bibitem{Bekenstein}  J. D. Bekenstein, Phys. Rev. D, {\bf 7}, 2333 (1973).

\bibitem{Hawking}  S. Hawking, Nature, {\bf 248}, 30 (1974).

\bibitem{Horowitz}  G. T. Horowitz, gr-qc/9704072.

\bibitem{Volovik/edge} G. E. Volovik, cond-mat/9709159.

\bibitem{KopninVolovik1995}  N. B. Kopnin, G. E. Volovik, and \"U. Parts,
Europhys. Lett. {\bf 32}, 651 (1995).

\bibitem{Wald}  R. M. Wald, gr-qc/9704008.

\bibitem{Bekenstein97}  J. D. Bekenstein, gr-qc/9710076.

\end{references}
\end{document}